\documentclass[amsmath,aps,prd,preprint,superscriptaddress]{revtex4-1}

\usepackage{graphicx,color}
\usepackage{multirow}
\usepackage{times}
\usepackage{bm}
\usepackage{epstopdf}
\usepackage{enumerate}

\usepackage[normalem]{ulem}  

\newcommand{\comment}[1]{}
\renewcommand\sout{\bgroup \color{red} \ULdepth=-.5ex \ULset}

\begin{document}

\title{
The enhancement of $v_4$ in nuclear collisions at the highest densities\\
 signals a first order phase transition
}

\author{Yasushi Nara}
\affiliation{
Akita International University, Yuwa, Akita-city 010-1292, Japan}
\affiliation{Frankfurt Institute for Advanced Studies, 
D-60438 Frankfurt am Main, Germany}
\author{Jan Steinheimer}
\affiliation{Frankfurt Institute for Advanced Studies, 
D-60438 Frankfurt am Main, Germany}
\author{Horst Stoecker}
\affiliation{Frankfurt Institute for Advanced Studies, 
D-60438 Frankfurt am Main, Germany}
\affiliation{Institut f\"ur Theoretishe Physik,
 Johann Wolfgang Goethe Universit\"at, D-60438 Frankfurt am Main, Germany}
\affiliation{GSI Helmholtzzentrum f\"ur Schwerionenforschung GmbH, D-64291
Darmstadt, Germany}

\date{\today}

\begin{abstract}
The beam energy dependence of $v_4$
(the quadrupole moment of the transverse radial flow)
is sensitive to the nuclear equation of state (EoS)
in mid-central Au + Au collisions
at the energy range of $3 < \sqrt{s_{NN}} < 30$ GeV,
which is investigated within the hadronic transport model JAM.
Different equations of state, namely, a free hadron gas,
a first-order phase transition and a crossover are compared.
An enhancement of $v_4$ at $\sqrt{s_{{NN}}}\approx 6$ GeV
is predicted for an EoS with a first-order phase transition.
This enhanced $v_4$ flow is driven by both the enhancement of $v_2$ as well as
the positive contribution to $v_4$ from the squeeze-out of
spectator particles which turn into participants
due to the admixture of the strong collective flow
in the shocked, compressed nuclear matter.
\end{abstract}

\maketitle

The azimuthal distribution of particles emitted,
in high energy heavy-ion collisions, 
contains important information
about the bulk properties of strongly interacting matter%
~\cite{Stoecker:1980vf,Stoecker:1981pg,Buchwald:1984cp,%
Stoecker:1986ci,Hartnack:1994ce,
Ollitrault:1992,Danielewicz:2002pu,Stoecker:2004qu}.
The azimuthal momentum distribution of particles
can be expressed as a Fourier series~\cite{Voloshin96,Poskanzer58,Voloshin0809},
\begin{equation}
E \frac{d^3N}{d^3p}  = \frac{1}{2\pi} \frac{d^2N}{p_Tdp_Tdy}
    \left(1 + \sum_{n=1}^\infty 2v_n \cos[( n (\phi - \Phi_n)]   \right)
\end{equation}
where $\phi$ is the azimuthal angle with respect to
the event plane $\Phi_n$, which is estimated experimentally
in various  ways.
The harmonic flow coefficients
\begin{equation}
  v_n = \langle \cos( n[\phi-\Phi_n]) \rangle
\end{equation}
measure the strength of the system response to the initial coordinate
space anisotropy and fluctuations in the collision zone.

Anisotropic flow is generated by the participant pressure%
~\cite{Stoecker:1980vf,Stoecker:1981pg}
during the early stages of the collisions,
therefore, is considered a sensitive messenger of
the equation of state (EoS)%
~\cite{Stoecker:1980vf,Stoecker:1981pg,Buchwald:1984cp,%
Stoecker:1986ci,Hartnack:1994ce,
Ollitrault:1992,Danielewicz:2002pu,Stoecker:2004qu}.
A large elliptic flow has been observed in RHIC and LHC experiments,
and is in good agreement with hydrodynamical simulations%
~\cite{Heinz:2013th,Gale:2013da,Huovinen:2013wma,Hirano:2012kj,%
Jeon:2015dfa,Jaiswal:2016hex,Romatschke:2017ejr}.
Hydrodynamical predictions revealed that the study of $v_4$ contains
 important information about the collision dynamics%
~\cite{Kolb:1999it,Kolb:2003zi,Borghini:2005kd,Gombeaud:2009ye,Luzum:2010ae}.
Recently, higher order coefficients $v_n$
have been measured at RHIC and LHC~\cite{Adams:2003zg,Acharya:2018zuq,%
Acharya:2018lmh}.\\

To investigate the phase structure of QCD, both the beam energy-,
centrality-, and system size- dependence are studied to access the different
regions of $T-\mu_B$ phase diagram~\cite{cbmbook}.
In particular, the search for a first-order phase transition and the critical
end point at high baryon density is a challenging goal  of
high energy heavy ion collisions~\cite{DHRischke2004}.

At lower beam energies ($\sqrt{s_{NN}}<10$ GeV),
the strength of the elliptic flow is determined
by the interplay between out-of-plane (squeeze-out)
and in-plane emission~\cite{Stoecker:1986ci,Sorge1997}.
In a previous work we predicted
a first-order phase transition~\cite{Chen:2017cjw,Nara:2017qcg}
will cause an enhancement of the elliptic flow $v_2$
as function of the beam energy
by the suppression of the squeeze-out due to
the softening of EoS~\cite{Zhang:2018wlk}.

Does this enhancement of $v_2$ suggest that $v_4$ is also enhanced
in the vicinity of a first-order phase transition?
This letter presents the beam energy dependence of $v_4$
as calculated with the microscopic transport model JAM~\cite{JAMorg},
using the modified scattering style method~\cite{Nara:2016phs,Nara:2016hbg}
and confirms our conjecture.
In JAM, particle production is modeled by the excitations of hadronic
resonances and strings, and their decays
in a similar way as in the RQMD and UrQMD models%
~\cite{Sorge:1995dp,UrQMD1,UrQMD2}.
Secondary products are allowed to scatter again, which generates 
collective effects within our approach.
In the standard cascade version of the model,
one usually chooses the azimuthal scattering angle randomly for any two-body scattering.
(The effects of a preserved two-body reaction plane have been studied 
in Ref.~\cite{Kahana:1994be}).
Thus, cascade simulations yield the free-hadronic gas EoS in equilibrium,
as then two-body scatterings, on average, do not generate additional pressure.
In our approach, the pressure of the system is controlled by changing
the scattering style in the two-body collision terms.
It is well known that an attractive orbit reduces the pressure,
while repulsive orbit enhances the pressure%
~\cite{Gyulassy:1981nq,Danielewicz:1995ay}.
Thus, the pressure is controlled by appropriately choosing
the azimuthal angle in the two-body scatterings.
Specifically,
the pressure difference from the free streaming pressure
$\Delta P$ is obtained by the following constraints:~\cite{Sorge:1998mk}:
\begin{equation}
\Delta P = \frac{\rho}{3(\delta\tau_i + \delta\tau_j)}
 (\bm{p}'_i-\bm{p}_i)\cdot (\bm{r}_i-\bm{r}_j) \,,
  \label{eq:eosmod}
\end{equation}
where $\rho$ is the local particle density and $\delta\tau_i$
is the proper time interval of the $i$-th particle
between successive collisions, ($\bm{p}'_i-\bm{p}_i$)
is the momentum change and $\bm{r}_i$ is the coordinate of
the $i$-th particle.
Momenta and coordinates in Eq.~(\ref{eq:eosmod}) refer to
the values in the c.m. frame of the respective binary collisions.
We had demonstrated that a given EoS can be simulated
by choosing the azimuthal angle according to
the constraint in ~Eq.(\ref{eq:eosmod})
in the two-body scattering process~\cite{Nara:2016hbg}.
We note that the total cross section and scattering angle of the
two-body scattering are not changed by this method;
the only modification is the choice of the azimuthal angle.

In this work, we use the same EoS as developed and used
in Ref.~\cite{Nara:2016hbg}
to simulate both the conjectured first-order phase transition (1OPT) and
also the alternative crossover transition (X-over).
The EoS with a first-order phase transition (EOS-Q)%
~\cite{Kolb:1999it,Sollfrank:1996hd}
is constructed by matching a free, massless quark-gluon phase
with the bag constant $B^{1/4}=220$ MeV with the hadron gas EoS.
In the hadronic gas phase, hadron resonances with mass up to 2 GeV
are included, with a repulsive, baryon density $\rho_B$ 
dependent mean field potential
$V(\rho_B)=\frac{1}{2}K\rho_B^2$, with $K=0.45$ GeV fm$^3$.
For the crossover EoS, we use the chiral model EoS from Ref.~\cite{CMEOS},
where the EoS at vanishing and
at finite baryon density is consistent with
a smooth crossover transition, i.e. 
this EoS is consistent with recent lattice QCD results.

For all presented results we compute $v_4$ with respect to the reaction plane:
$\Phi_n=\Phi_{RP}$,
where $\Phi_{RP}$ is the reaction plane angle of the collision.
As usual, the reaction plane anisotropies in the even-order Fourier coefficients
are in good agreement with the anisotropies
taken with respect to the event plane,
while odd-order Fourier coefficients are generated by event-by-event
fluctuations.

\begin{figure}[tbh]
\includegraphics[width=12.0cm]{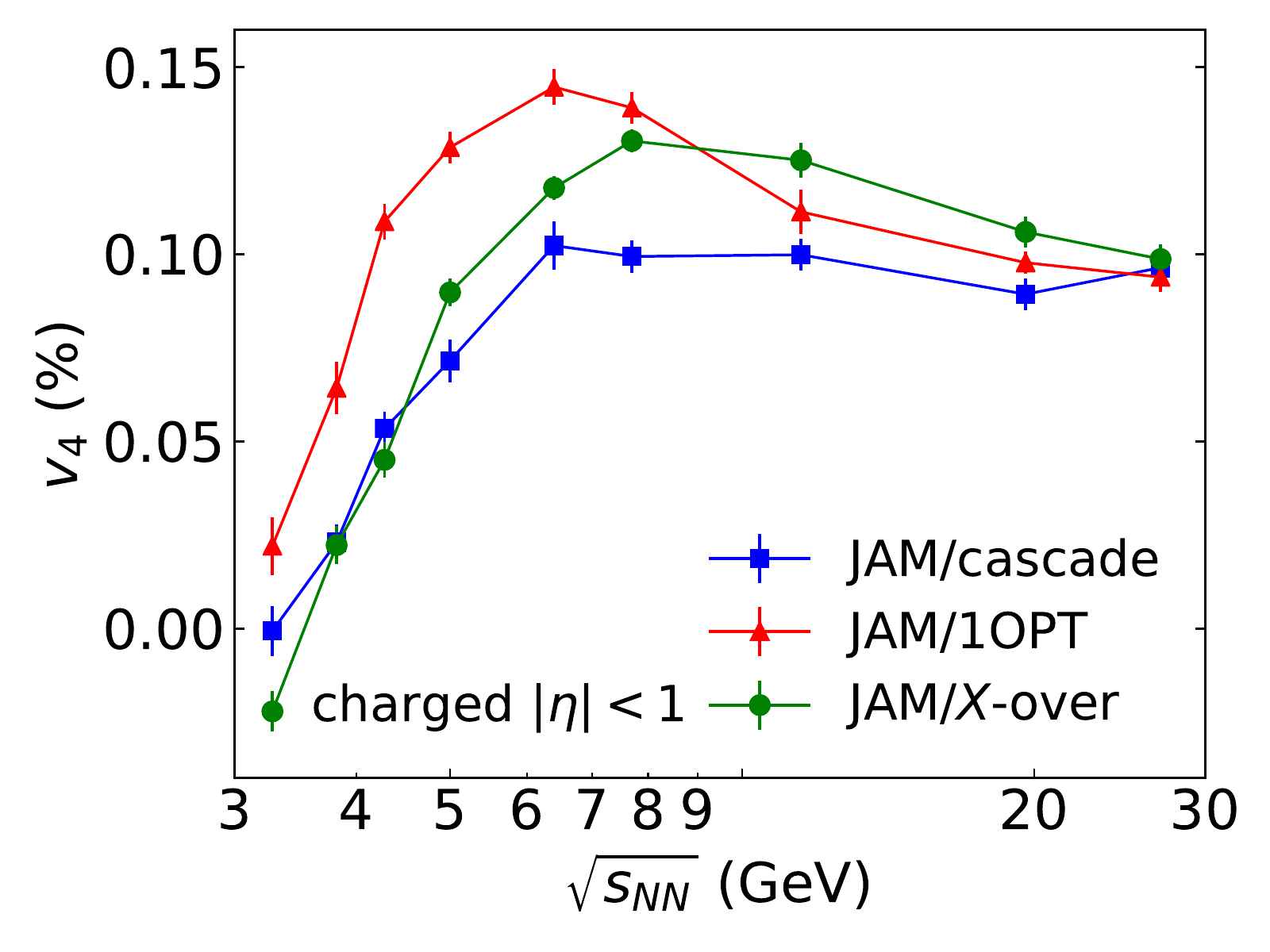}
\caption{Beam energy dependence of the $v_4$
for charged hadrons at $|\eta|<1.0$
in mid-central Au+Au collisions ($4.6\leq b\leq9.4$ fm)
from the JAM cascade mode (squares),
JAM with first-order EoS (triangles),
and crossover EoS (circles).
}
\label{fig:v4exfunc}
\end{figure}

Figure~\ref{fig:v4exfunc} shows the beam energy dependence of
$v_4$, for charged particles at mid-rapidity $|\eta|<1.0$
in mid-central Au + Au collisions
from the JAM model with the cascade mode,
JAM with the first-order EoS (JAM/1OPT),
and a crossover EoS (JAM/X-over).
The effects of our three different EoS on the $v_4$ at higher beam energy
$\sqrt{s_{NN}}>10$ GeV are quite similar,
in contrast to the high baryon density,
i.e. at $\sqrt{s_{NN}}<10$ GeV,
where the effect of the EoS is very strong.
The cascade mode results do not show any clear maximum or bump
in the beam dependence of the $v_4$.
The calculations using an EoS with a first-order phase transition
and those with
a crossover transition
exhibit an enhancement of $v_4$ relative to the cascade result
at 5 GeV, a factor of two for the 1OPT case,
and an inversion of sign of $v_4$ at 3 GeV for the X-over case.
JAM/1OPT shows a strong bump around
the beam energy of $\sqrt{s_{NN}}\approx 6$ GeV.
A similar enhancement was observed
in the case of $v_2$ for the 1OPT mode~\cite{Nara:2017qcg}.

\begin{figure*}[tbh]
\includegraphics[width=7.2cm]{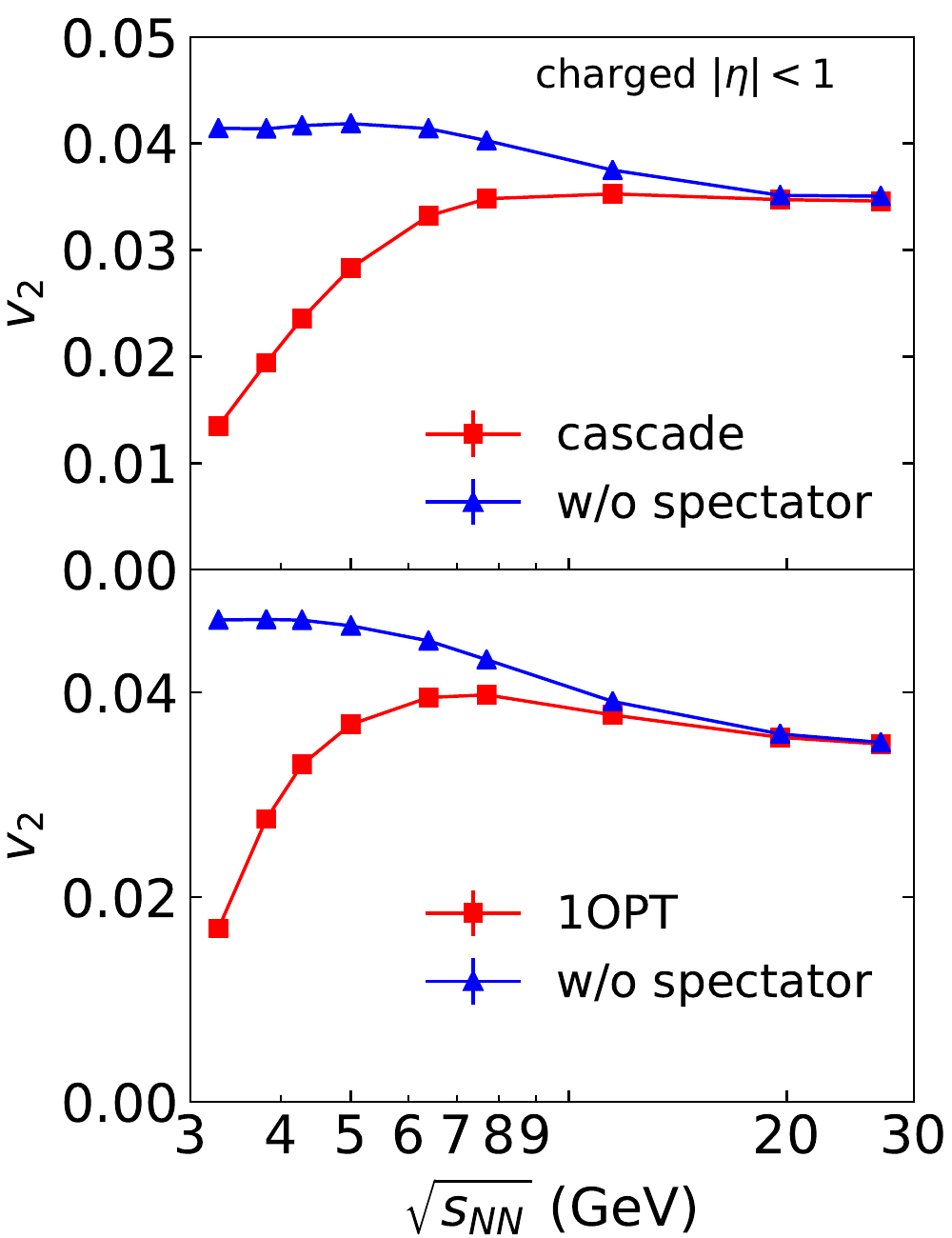}
\includegraphics[width=7.2cm]{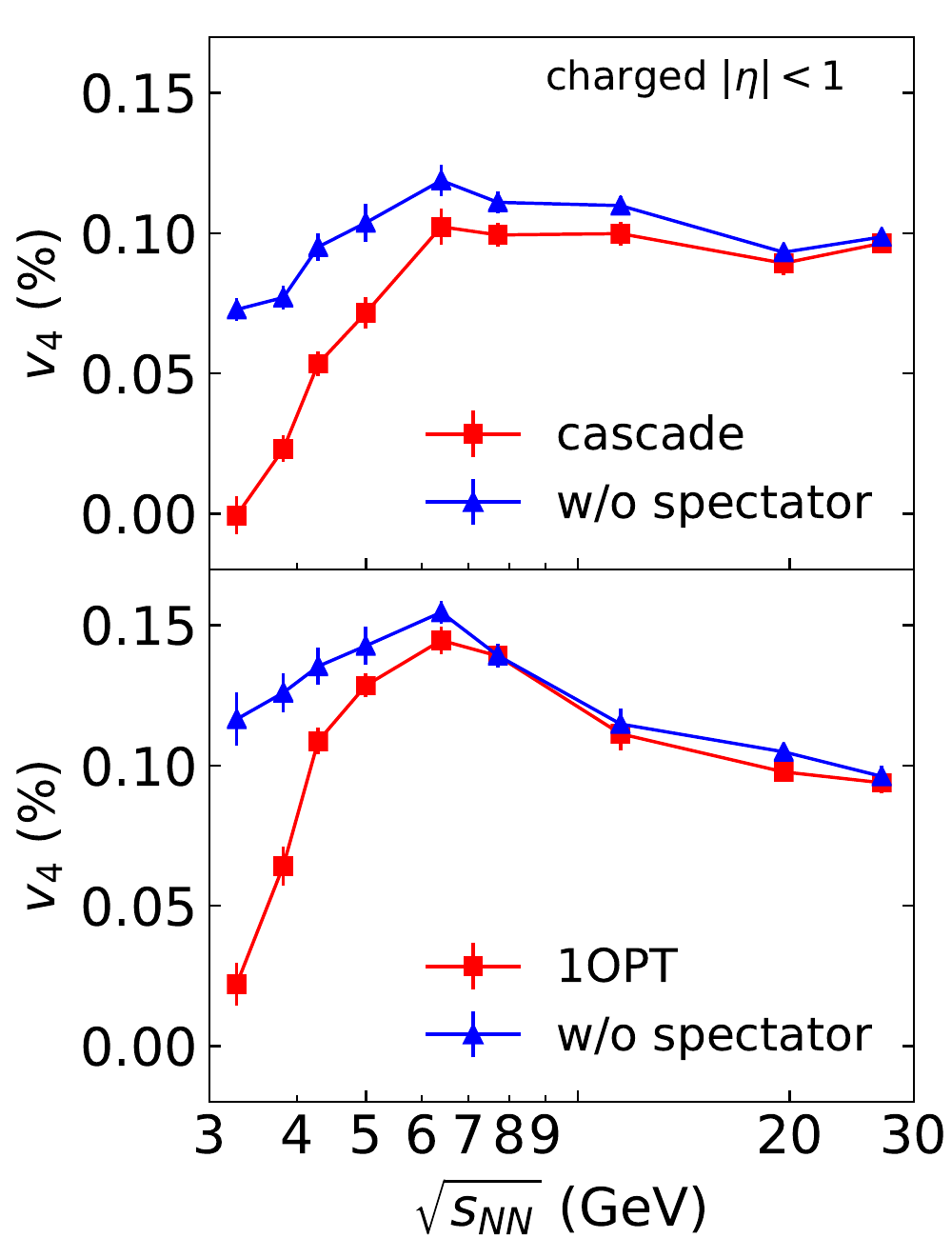}
\caption{Beam energy dependence of the $v_2$- (left panel)
and the $v_4$ (right panel) coefficients
in mid-central Au + Au collisions,
with- and without spectator interactions,
are compared to the different EoS in JAM simulations.
}
\label{fig:v2spec}
\end{figure*}

To understand the collision dynamics which enhances
both the $v_2$ and the $v_4$, 
we consider the effects of spectator interactions:
Out-of-plane emission (squeeze-out) is mainly driven by
the pressure release perpendicular to the spectator
plane, which yields the negative
$v_2=\left\langle \frac{p_x^2-p_y^2}{p_T^2}\right\rangle$
at lower beam energies. 
In the beam energy range of $3<\sqrt{s_{NN}}<10$ GeV,
the cancellation between
the in-plane flow ($p_x$) and the out-of-plane flow ($p_y$)
determines the final value of $v_2$.
Thus, if the spectator-matter interaction is neglected,
the elliptic flow is strongly positive.
To see the effects of spectator interactions on the flows quantitatively,
we perform the calculations in which interactions with 'spectator nucleons'
are disabled, where 'spectator nucleons' are defined as
the nucleons which are not in the list of initial collisions;
collisions of nucleons which are initially located
outside the overlap region of the two colliding nuclei
therefore are excluded in the calculations without spectator matter.

Figure~\ref{fig:v2spec} compares the calculations of flow with and
without 'spectator nucleons'.
If the EoS with the first-order phase transition is employed,
the effect of spectator shadowing is smaller than in the cascade mode,
as the pressure is significantly smaller and, hence,
the acceleration of the stopped matter is less for 
this softest equation of state -- then
the system remains in this low pressure region for a long time
reached in the system.
This is the origin of the enhancement of $v_2$ if there is a first-order
phase transition. 

In the following we will discuss the effects of the spectator matter on $v_4$:
The elliptic flow $v_2$ is positive in the case of stronger in-plane emission,
see the left-hand side of Fig.~\ref{fig:v2spec},
while $v_2$ is negative for predominant out-of-plane emission
at $\sqrt{s_{NN}}<3$ GeV.
On the other hand, $v_4$ is positive, and large for both, in-plane and
out-of-plane emission.
Thus, spectator shadowing will enhance the $v_4$ value.
Thus, it is indeed seen in Fig.~\ref{fig:v2spec} (right panel)
if the spectator interactions are neglected,
$v_4$ is not suppressed, up to $\sqrt{s_{NN}} = 6$ GeV.
In the case of a first-order phase transition, $v_4$ does increase,
both with and without spectator interactions.
At the lower beam energies, $v_4$ decreases for both calculations,
with and without spectator interactions, in contrast to $v_2$,
which increases at lower energies if the spectator interactions are
neglected.
Here, particle emission is not so strongly
directed to the in-plane direction, which decreases
the $v_4$ at lower beam energies $\sqrt{s_{NN}}<5$ GeV. 
One should note that at even lower beam energies
$\sqrt{s_{\mathrm{NN}}} \le 4$ GeV
the effects of nuclear potentials need to be taken into account
for quantitative predictions on the $v_4$.

\begin{figure}[t]
\includegraphics[width=8.5cm]{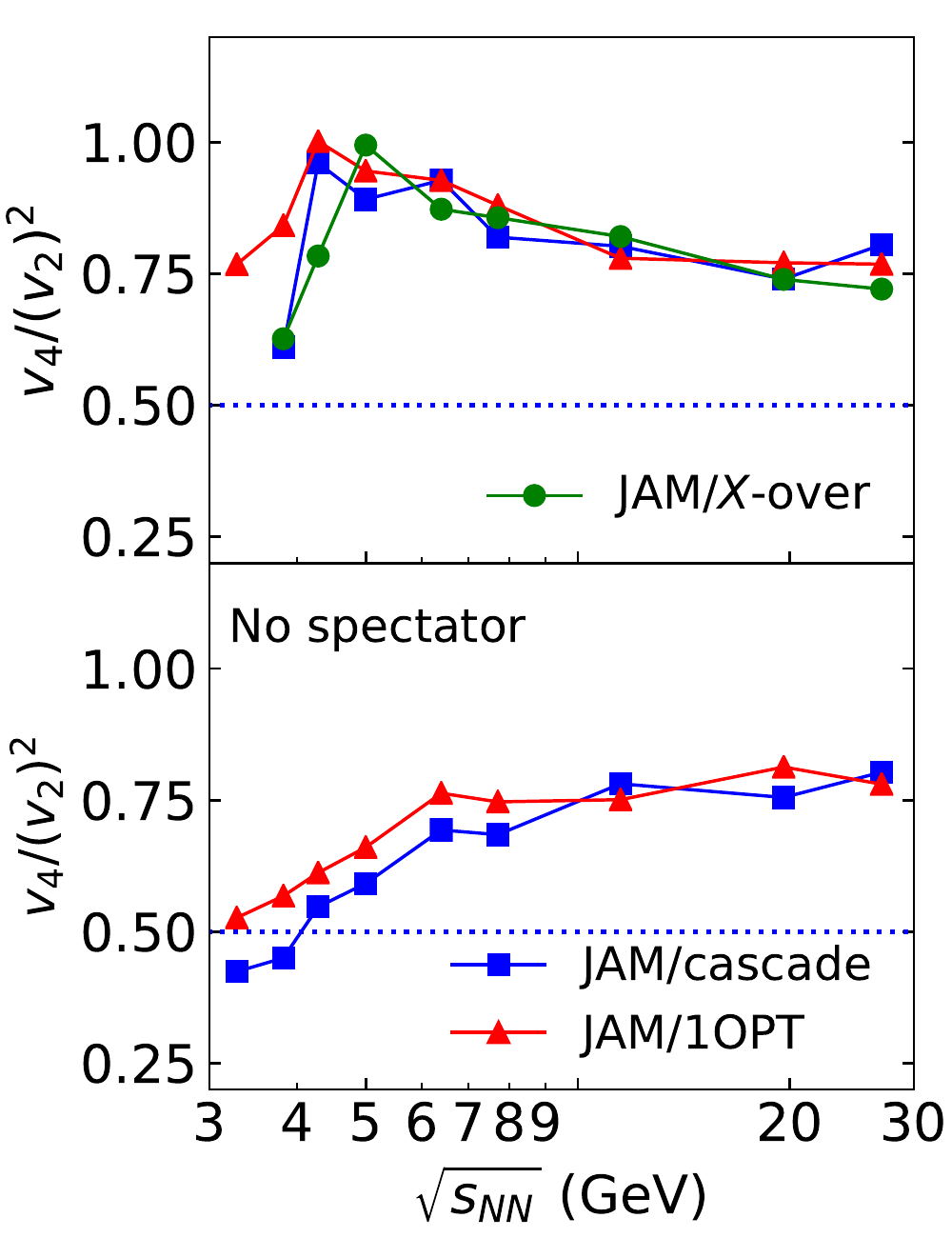}
\caption{Same as Fig.~\ref{fig:v4exfunc}, but
for the $v_4/(v_2)^2$ ratio of charged particles
with (upper panel) and without spectator matter interactions
(lower panel).
}
\label{fig:v2v4spec}
\end{figure}

The harmonic $v_4$ is generated both by the intrinsic $v_2$
and by the forth-order moment of the collective flow%
~\cite{Borghini:2005kd,Gombeaud:2009ye,Luzum:2010ae}.
Within ideal fluid dynamics (and without any fluctuations),
the elliptic flow contribution to $v_4$ is simply given by
$v_4 =0.5(v_2)^2$~\cite{Borghini:2005kd,Gombeaud:2009ye,Luzum:2010ae}.
Hence, the ratio $v_4/(v_2)^2$ contains valuable information
about the intrinsic collision dynamics.
Experimental data show that $v_4$ is about double the ideal hydro values,
$v_4 \approx (v_2)^2$ at RHIC%
~\cite{Adams:2003zg,Masui:2006,Abelev:2007,Huang:2008}.
Note that the PHSD results show a fourfold higher value,
$v_4/(v_2)^2\approx 2$~\cite{Konchakovski:2012yg},
for a wide range of beam energies in min-bias Au + Au collisions.

Figure~\ref{fig:v2v4spec} shows the beam energy dependence of
the $v_4/(v_2)^2$ ratio,
stands close to 0.75 at 3 GeV, then rise,and flatters to 
a constant value of 0.75 at
beam energies of $\sqrt{s_{NN}}>10$ GeV,
with  a slight increase around 6 GeV.
Calculations where spectator matter interactions are neglected
yield smaller values $v_4/(v_2)^2 \sim 0.5$ at moderate energies,
but also approach 0.75 at $\sim6$ GeV, and above.
This indicates that the $v_4$ is dominated by the $v_2$ component
as without spectator shadowing there exists no squeeze-out effect.
Actually, the beam energy dependence of $v_2$ exhibits
a similar dependence as $v_4$, in the simulations without spectator matter,
as can be seen in Fig.~\ref{fig:v2spec}.

In summary, we have studied the beam energy dependence of
the fourth harmonics $v_4$ for charged particles in mid-central Au + Au
collisions at $3 < \sqrt{s_{NN}} < 30$ GeV.
An enhancement of $v_4$ around beam energies of 6 GeV is predicted
if and only if a first-order phase transition is present-
hence, this can serve as a clean signal.
The enhancement of $v_2$ is caused by the weaker squeeze-out effects
exerted by the spectator matter, due to the soft EoS.
An enhancement of $v_4$ comes from the enhancement of $v_2$ itself
as well as from
the positive contributions from the squeeze-out.

Predicted $v_4$ signal can be studied experimentally at
future experiments such as RHIC-BESII~\cite{BESII},
FAIR~\cite{FAIR,Sturm:2010yit}
NICA~\cite{NICA}, and J-PARC-HI~\cite{HSakoNPA2016},
which offer the best opportunities to
explore the compressed baryonic matter, and
reveal the phase structure of QCD.

\begin{acknowledgments}
Y. N. thanks the team of the Frankfurt Institute of Advanced Studies
where part of this work was done for their splendid hospitality.
This work was supported in part by the
Grants-in-Aid for Scientific Research from JSPS
(JP17K05448).
H. St. appreciates the generous endowment of the 
 Judah M. Eisenberg Laureatus professorship.
Computational resources have been provided by the Center for Scientific
Computing (CSC) at the J. W. Goethe-University, Frankfurt,
and GSI, Darmstadt.
J. S. appreciates the support of the SAMSON AG and the C.W. Fueck-Stiftungs Prize 2018.
Computational resources have been provided by the Center for Scientific
Computing (CSC) at the J. W. Goethe-University, Frankfurt,
and GSI, Darmstadt.

\end{acknowledgments}

\end{document}